\documentclass{iopart}
\usepackage{amsmath,graphicx,bbm,iopams,mathrsfs,psfrag}
\newcommand{\bra}[1]{\langle #1 \vert}
\newcommand{\ket}[1]{ \vert#1\rangle }
\newcommand{\bmsigma}{\boldsymbol \sigma}

\begin{document}
\title{Bayesian estimation of one-parameter qubit gates}
\author{Berihu Teklu}
\address{Dipartimento di Fisica, Universit\`a di Milano,
I-20133 Milano, Italy}
\author{Stefano Olivares}
\address{ CNISM, UdR Milano Universit\`a, I-20133 Milano, Italy
\\ Dipartimento di Fisica, Universit\`a di Milano, I-20133 Milano, Italy}
\author{Matteo G. A. Paris}
\address{
Dipartimento di Fisica, Universit\`a di Milano, I-20133 Milano, Italy
\\CNISM, UdR Milano Universit\`a, I-20133 Milano, Italy
\\Institute for Scientific Interchange Foundation, I-10133 Torino, Italy}
\date{\today}
\begin{abstract}
We address estimation of one-parameter unitary gates for qubit systems
and seek for optimal probes and measurements. Single- and two-qubit
probes are analyzed in details focusing on precision and stability of
the estimation procedure. Bayesian inference is employed and compared
with the ultimate quantum limits to precision, taking into account the
biased nature of Bayes estimator in the non asymptotic regime. Besides,
through the evaluation of the asymptotic {\em a posteriori} distribution
for the gate parameter and the comparison with the results of Monte
Carlo simulated experiments, we show that asymptotic optimality of Bayes
estimator is actually achieved after a limited number of runs.  The
robustness of the estimation procedure against fluctuations of the
measurement settings is investigated and the use of entanglement to
improve the overall stability of the estimation scheme is also analyzed
in some details.
\end{abstract}
\section{Introduction}\label{s:intro}
Let us consider a system prepared in a known quantum state which enters
an apparatus performing an operation on the state. The evolution imposed
by the apparatus depends on the value of some parameters and the
experimenter is interested in the estimation of those parameters. 
A natural strategy to obtain the parameter is to detect the state at 
the output and infer the value of the
parameters from the global sample coming from a number of repeated
measurements. The optimization of this strategy, {\em i.e.} the choice
of the best probe, measurement and data processing, is the subject of quantum
parameter estimation, which is a relevant subject in the quantum
characterization of states and operations \cite{Hel76,Hol82}.  The
operation on the state may be unitary or not \cite{ChuNie,PoyCirZol97}
and may depend on one or more unknown parameters, which, in turn, may
correspond to quantities that are not directly observable.
The parameters of interest may be the amplitude of the carrier
signal, the position and orientation of an object, or the strength of an
external fields. Communications, image analysis and precision metrology 
provide relevant examples. 
In the simplest scenario, a parameter estimation problem
consists in the determination of the value of the interaction parameter
$\theta$ for unitaries of the form $U_{\theta} =\lbrace -i \theta
G\rbrace$ where $G$ is a Hamiltonian operator that generates the
transformation. 
\par
Generally speaking, quantum estimation is concerned with the problem of
finding optimal ways to estimate quantum states and processes. In turn,
it has recently attracted much interest in quantum information
\cite{Braunstein1994, Hol2004,Gill,Barnd,Fuji1995} as a tool for
characterization of signal and gates at the quantum level
\cite{qsm,Braunstein1,Braunstein2,Braunstein3,gentomo,mkq,Boixo,EKnill,Giovannetti}.
The canonical way to address estimation of states and operation is by
quantum tomography, (see \cite{LNP649} for a review) {\em i.e} by
measuring a complete set (a {\em quorom} \cite{gentomo}) of observables,
which allows or the complete characterization. For single- and two-qubit
systems this involves the measurement of Pauli matrices and has been
realized for polarization qubits \cite{kw2,qbmi}.  Process Tomography,
{\em i.e} the reconstruction of quantum operations \cite{op1,op2,op3}, 
is itself critical for verifying the actions of quantum logic gates 
\cite{op4} and characterizing decoherence processes \cite{op5}.
On the other hand,
there are many situations where the full tomography of signals and
devices is not needed, either because the focus is on specific features
of the transformation, or the dynamics is partially known. In this cases
the relevant point is to find an optimal and stable way to achieve
quantum characterization by parameter estimation. For this reason, in
this paper we address estimation of one-parameter unitary gates for
qubit systems, {\em i.e.} transformation of the form $U_{\theta}
=\lbrace -i \theta G\rbrace$ where $G$ is a combination of Pauli
operators and $\theta$ is the parameter of interest. We consider the
gate probed either by one-qubit and two-qubit states and compare the
performances of standard measurements with the ultimate quantum limit to
precision (accuracy) of estimation.  As we will see, ultimate bounds are
determined by the initial quantum state of the probe, the type of
interaction and the readout measurements that is used to extract
information from the probe. In particular, we are going to assess the
performances of Bayes estimators, which themselves play a central role
in many signal processing problems \cite{VTB}. 
\par
The precision (variance) of any unbiased estimator of a parameter
$\theta$ is limited by the Cram\'er-Rao bound (CR), given by the inverse
of the Fisher information \cite{Gill,Cramer,Rao,Poor,qi,H1,H2}.  Bayes
estimators are known to be asymptotically unbiased and, in turn, to
saturate CR asymptotically. For measurements that are related to the
unknown $\theta$ through a linear Gaussian model, the maximum likelihood
estimate of $\theta$ also achieves the CR. Furthermore, when $\theta$ is
estimated from independent identically distributed (iid) measurements,
under suitable regularity assumptions on the probability density, the
maximum likelihood estimator is asymptotically unbiased and achieves the
CR \cite{Rao,Leh}. On the other hand, being interested in realistic
measurement schemes, here we consider estimation procedure based on a
limited number of measurements. As a consequence, we have to take into
account the biased nature of Bayes estimators. The variance of any
estimator with a given bias is bounded by the biased CR
\cite{VTB1,Schi}, which is an extension of the CR taking into account
the {\em a priori} distribution of the parameter of interest. In turn,
it is a fundamental rule of estimation theory that the use of prior
knowledge leads to a more accurate estimator. 
\par
In this paper we address the estimation of the interaction parameter of
unitary qubit transformations. We derive ultimate quantum limits to
precision and assess performances of Bayesian estimators \cite{zd,pz}. 
In particular,
we focus our attention on measurement schemes as those in Fig.
\ref{fig:qubit} and Fig. \ref{fig:bell}, where a single-qubit gate is
probed by single- and two-qubit probes, respectively. We evaluate the a
posteriori distribution for the gate parameter, derive the ultimate
bound on precision, and compare the asymptotic performances of Bayes
estimator to that of Monte Carlo simulated experiments, thus showing
that asymptotic optimality is achieved after a limited number of runs.
The paper is structured as follows. In Section \ref{s:estimation} we
introduce notation and derive the a posteriori distribution, also
discussing the Bayesian version of the CRB. In Section \ref{s:single} we
discuss limits to precision in estimating unitary gates for qubit
systems.  A comparison between single- and two-qubit entangled probes shows
that entanglement improves the overall stability of the estimation
procedure. We also compare the asymptotic a posteriori
distribution for the gate parameter to the results of the Monte Carlo
simulated experiments. Section \ref{s:remarks} closes the paper with
some concluding remarks.
\section{Parameter estimation of one-parameter qubit gates}
\label{s:estimation}
A generic unitary transformation acting on a qubit state can be written
as $U({\boldsymbol\theta}) = \exp\lbrace - \frac{i}{2}
{\boldsymbol\theta}\cdot{\bmsigma}\rbrace$,
where ${\bmsigma}=\lbrace\bmsigma_1,\bmsigma_2,\bmsigma_3\rbrace$ is
the vector of Pauli matrices and ${\boldsymbol\theta}$ is a vector 
describing
the transformation. In the following we will assume that ${\boldsymbol\theta}$
depends on a single parameter and refer to it as to $\theta$.  The
simplest scheme for the estimation of $\theta$ \cite{Hel76} consists in
choosing a {\em probe} pure qubit state $\ket{\psi_0}$ undergoing the
transformation and a measurement onto the evolved state
$\ket{\psi_\theta} \equiv U({\theta})\ket{\psi_0}$. Here we assume that
the measurement can be represented by the two {\em projectors} $\Pi_0$
and $\Pi_1$, so that the {\em conditional} probabilities to obtain the
outcomes ``0'' or ``1'' ({\em i.e.} given $\theta$) are $P(j|\theta) =
\bra{\psi_\theta}\Pi_j\ket{\psi_{\theta}}$, $j=0,1$.
After $M$ measurements on
equally prepared qubits we have the the sample $X = \{x_1,\ldots, x_M\}$
of outcomes, where the $x_k$'s can take the values ``0'' and ``1''. This
leads us to define the following sample probability or likelihood
function:
\begin{equation}
P(X|\theta) = \prod_{k=1}^{M} P(x_k|\theta).
\end{equation}
Our estimation problem is that of inferring the value of $\theta$ once
the sample of outcomes is assigned by the measurement; in other words,
we are interested in the conditional ({\em a posteriori}) probability
$P(\theta|X)$ of $\theta$ {\em given} the sample $X$. This can be easily
obtained by the Bayes theorem, which states that \cite{JMB,DM}
$P(\theta|X) P(\theta) = P(X|\theta) P(X)$ where $P(\theta)$ is the
prior probability and $P(X)$ is the overall (unconditional) probability
of the observed sample.  Hence, the {\em a posteriori} distribution 
may be written as 
\begin{equation}\label{cond:th}
P(\theta|X) = \frac{1}{N} \prod_{k=1}^{M} P(x_k|\theta)\,, \qquad
{N} = \int_{\Omega}\!\!\! d\theta \prod_{k=1}^{M} P(x_k|\theta)\,,
\end{equation}
where ${N}$ is the normalization and $\Omega$ is the set of 
possible values for $\theta$. 
From (\ref{cond:th}) we may evaluate the expected value of 
$\theta$ and the variance of the distribution
\begin{equation}
\overline{\theta} =
\int_{\Omega}\!\!\! d\theta \, \theta \, P(\theta|X),
\quad \quad
{\rm Var}[\theta] =
\int_{\Omega}\!\!\! d\theta \, (\theta-\overline{\theta})^2 \, 
P(\theta|X)\:.
\end{equation}
The mean (expected) value $\overline{\theta}$ of the a posteriori 
distribution is our Bayesian estimator.
\par
For a large number of measurements, $M \gg 1$, and assuming
that the true value of the parameter is $\theta^*$, the number of
times a factor $P(x|\theta)$, with $x = 0, 1$, appears in the product
(\ref{cond:th}) is approximately given by $P(x|\theta^*) M$. The 
asymptotic a posteriori distribution for the parameter $\theta$, conditioned
on the true value $\theta^*$ is thus given by \cite{Hradil1995,Hradil1996}
\begin{equation} \label{LF}
P_M(\theta |\theta^{*}) = \frac{1}{N}\prod_{s=0,1}
P(s|\theta)^{P(s|\theta^{*}) M}\,.
\end{equation}
Since $\sum_{h=0,1} P(h|\theta) = 1$ we have 
$ \partial_\theta P_M(\theta|\theta^{*})\left.
\vphantom{\frac{}{}}\right|_{\theta^{*}} =0$ and 
$\partial^2_\theta P_M(\theta|\theta^{*})\left.
\vphantom{\frac{}{}}\right|_{\theta^{*}} < 0$, {\em i.e} the distribution 
$P_M(\theta|\theta^*)$ has the desirable property of showing a maximum 
at the true value of the parameter, {\em i.e.} Bayesian estimator is 
asymptotically unbiased.
\par
According to the Laplace-Bernstein-von Mises theorem \cite{LeCam} the a posteriori 
distribution  of Eq.~(\ref{LF}) may be asymptotically approximated by a Gaussian 
with variance given by $\sigma^2 = \left[M G(\theta^*)\right]^{-1}$ 
where we have introduced the {\em Fisher information} \cite{qi}:
\begin{equation}
G(\theta) = \sum_{s=0,1} \frac{1}{P(s|\theta)}\,
\left(\frac{d P(s|\theta)}{d\theta}\right)^{2}
\label{deffisher}
\end{equation}
The asymptotic a posteriori distribution is thus 
completely characterized by its variance, or equivalently by 
the Fisher information which itself gives a lower bound to 
the variance of any unbiased estimator 
$\hat\theta(X)$ via the Cram\'er-Rao inequality \cite{Gill,qi,H1,H2}:
\begin{equation}
{\rm Var}_\theta[\hat\theta] \ge \frac1{M G(\theta)}\:. \label{cramerrao}
\end{equation}
Any estimator saturating the inequality (\ref{cramerrao}) is referred
to as an {\em efficient} estimator. The relation  $\sigma^2 = \left[
M G(\theta^*)\right]^{-1}$ thus says that Bayesian estimator is 
asymptotically efficient, whereas this conclusion does not hold for 
finite $M$. For finite $M$ Bayes estimator is biased and we have to 
generalize Eq.~(\ref{cramerrao}) to take into account the bias. To this 
aim one considers the conditional expectation of the error $B(\theta)=
\int dx\, [\hat\theta(x) -\theta]P(x,\theta)$, where $P(x,\theta)=P(x|\theta)
P(\theta)$ is the joint probability of the data and the parameter. Of
course, for unbiased estimators $B(\theta)=0$. Starting from the definition of 
$B(\theta)$ one derives the so-called van Trees inequality \cite{VTB1} for 
the mean squared error
\begin{equation}\label{VT:ineq}
\overline{{\rm Var[\hat \theta(x)-\theta]}} 
= \int d\theta\,P(\theta) \int dx\, [\hat\theta(x)-\theta]^2 P(x|\theta) 
\geqslant
\frac1{H_M(\theta)}\:,
\end{equation}
where we introduced the generalized Fisher information 
$H_M(\theta)={F(\theta)+M\,G(\theta)}$, $G$ being the Fisher information of 
Eq. (\ref{deffisher}), $M$ the number of repeated measurements, and $F$ 
the Fisher information of the prior, {\em i.e.} $F(\theta)=\int d\theta\,
\left[\partial_\theta \log P(\theta) \right]^{2}P(\theta)$.
Eq.~(\ref{VT:ineq}) takes into account the information due to the prior 
and thus gives a lower bound than the CR one, which is anyway achieved 
for $M\gg 1$. On the other hand, one may show that \cite{Schi}:
\begin{align}
H_M(\theta) &= \int\!\! dX\, P(X) \int\!\! d\theta\,
\left[ \partial_\theta \log P(\theta | X)\right]^2 P(\theta | X)
\nonumber \\
& \stackrel{M\gg 1}{=} F_M(\theta|\theta^*) \equiv \int\!\! d\theta
P_M(\theta|\theta^* ) \left[ \partial_\theta \log P_M(\theta|\theta^* )
\right]^2\,,
\end{align}
where $P(\theta | X)$ is the Bayesian probability distribution
of Eq. (\ref{cond:th}) and $P_M(\theta|\theta^* )$ its asymptotic
expression of Eq. (\ref{LF}). Thus, Eq.~(\ref{VT:ineq}) represents the
Bayesian counterpart of the CR.
\section{Bayesian estimation of one-parameter qubit gates}
\label{s:single}
In this section, without loss of generality, we address the case in which 
$\boldsymbol{\theta}=(0,0,\theta)$, $\theta\in[0, \pi]$, i.e., $U_3(\theta) =
\exp(-\frac{i}{2}\theta\bmsigma_3)$. We first consider the gate probed 
by a single-qubit state and jointly optimize the probe and the
measurement and then address the use of entanglement, showing that it
may be useful to improve the overall stability of the estimation
procedure. 
\subsection{Estimation via sinqle-qubit probes}
\begin{figure}[h]
\includegraphics[width=0.6\textwidth]{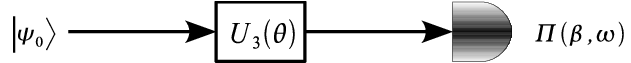}
\caption{Estimation of qubit gates. The state $\lbrace\rho_{0}\rbrace$
is prepared, then enters the gate described by the unitary $U_3(\theta)$
with unknown $\theta$, and finally is detected by a detector described
by a two-value POVM. \label{fig:qubit}}
\end{figure}

In Fig.~\ref{fig:qubit} we schematically depict the estimation 
procedure: a pure state $\ket{\psi_0}$ undergoes the unitary
transformation $U_3(\theta)$ and, then, is measured by means of a
projective two-outcome device. Our aim is that of optimizing 
the estimation of $\theta$ by a suitable choice of both the probe state 
and the projective measurement. 
\par
\begin{figure}[h!]
\psfrag{thst}{$\overline {\theta}/\theta^{*}$}
\psfrag{M}{$M$}
\psfrag{V}{${\rm Var}[\theta] {H_M}$}
\psfrag{300}[]{300} \psfrag{500}[]{500} \psfrag{700}[]{700}
\psfrag{1.0000}[]{1.000} \psfrag{1.0005}[]{}
\psfrag{1.0010}[]{1.001} \psfrag{1.0015}[]{}
\psfrag{1.0020}[]{1.002} \psfrag{1.0025}[]{} \psfrag{1.0030}[]{1.003}
\includegraphics[width=0.41\textwidth]{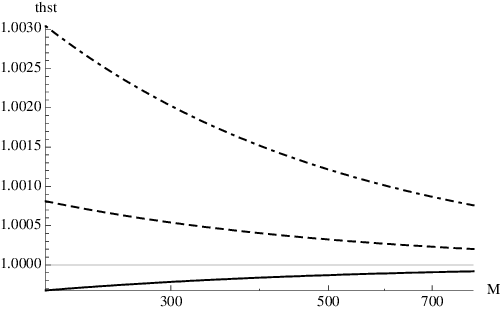}
\psfrag{M}{$M$} \psfrag{20}[]{20} \psfrag{40}[]{40} \psfrag{60}[]{60}
\psfrag{80}[]{80} \psfrag{100}[]{100}
\psfrag{1.}[]{1.} \psfrag{1.01}[]{1.01} \psfrag{1.02}[]{1.02}
\psfrag{1.03}[]{1.03} \psfrag{1.04}[]{1.04} \psfrag{1.05}[]{1.05}
$\qquad$\includegraphics[width=0.41\textwidth]{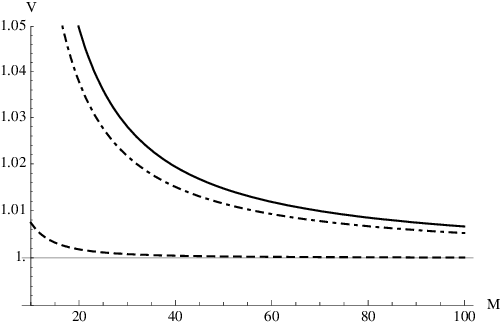}
\caption{LogLinear plot of $\overline{\theta}/\theta^{*}$ (left) 
and linear plot of ${\rm Var}[\theta]{H_M}$ (right) as a function 
of the number of measurements $M$ for the estimation of the unitary
$U_3(\theta)$ and for different values of the true parameter. For both 
plots the dotted-dashed line is for $\theta^{*}=0.8$, the dashed line 
is for $\theta^{*}=1.2$ and the solid line is for $\theta^{*}=1.8$.}
\label{f:uni}
\end{figure}

Upon writing the pure qubit state in the standard way 
\begin{equation}
|\psi_{0}\rangle=\cos \frac\alpha {2}|0\rangle
+  e^{i \phi} \sin \frac \alpha {2}|1\rangle\:,
\end{equation}
where $|0\rangle$ and $|1\rangle$ are the two eigenvectors of
$\bmsigma_3$, the parameters $\alpha\in[0,\pi]$ and $ \phi\in[0,2\pi]$
uniquely determine $|\psi_{0}\rangle$ and the evolution under the unitary
$U_3(\theta)$ is straightforward.
Next, we perform the measurement described by the two projectors 
\begin{align}
\Pi_0(\beta,\omega) =
\ket{\Psi(\beta,\omega)}\bra{\Psi(\beta,\omega)}\qquad
\Pi_1(\beta,\omega) = {\mathbbm 1} - \Pi_0(\beta,\omega),
\end{align}
where $|\Psi(\beta,\omega)\rangle=\cos \frac\beta {2}|0\rangle
+  e^{i \omega} \sin \frac \beta {2}|1\rangle$. The probabilities of 
the two outcomes are given by
$P(0|\theta) \equiv P_0(\alpha,\beta,\phi,\omega,\theta) = 
|\langle 0|U_{\theta}|\psi_{0}\rangle|^{2}$ {\em i.e}
\begin{align}
\label{prob1}
P(0|\theta) &=  \frac12 \left[1 + \cos\alpha\cos\beta - \cos
(\phi-\omega+\theta) \sin\alpha\sin\beta\right] \nonumber \\
P(1|\theta) &= 1-P(0|\theta)\:.
\end{align}
The Bayesian a posteriori distribution of Eq.  (\ref{cond:th}) may be written as 
\begin{align}
\label{ge}
P(\theta|M) = \frac1N \: P(0|\theta)^{m_0}\:P(1|\theta)^{m_1}\:,
\end{align}
where $m_j$ is the number of measurements with outcomes 
$j=0,1$, $m_0+m_1=M$, in the observed sample.
For a large number of measurements $M\gg 1$, we can evaluate
$P_M(\theta|\theta^*)$ using Eq. (\ref{LF}) and, in turn, the
expectation $\overline\theta$ and the variance ${\rm Var}[\theta]$. 
Upon expanding on $\alpha$ and $\beta$ up to second
order one sees that ${\rm Var}[\theta]$ achieves its minimum for the
choice $\alpha = \beta = \pi/2$, independently on the value of $\phi$ and
$\omega$. This is confirmed by the evaluation of the corresponding
Fisher information $G (\theta)$, 
\begin{align}
G(\theta) = \frac{\sin^2 \alpha \sin^2 \beta
\sin^2(\theta+\phi-\omega)}{1-\cos^2 \alpha \cos^2 \beta -
\cos^2(\theta+\phi-\omega)\sin^2 \alpha \sin^2 \beta }\:, 
\end{align}
which achieves its maximum $G=1$ for $\alpha = \beta = \pi/2$, 
independently on the value of $\phi$ and $\omega$. Using this
results we have:
\begin{align}\label{ga}
P_{M}(\theta|\theta^{*})=\frac1{N}
\exp\left[M\left(\cos^{2}\frac{\theta^{*}}2\log \cos^{2}\frac{\theta}2
+\sin^{2}\frac{\theta^{*}}2\log\sin^{2}\frac{\theta}2\right)\right],
\end{align}
and the variance saturates the van Trees inequality, thus confirming
that Bayes estimator is asymptotically efficient.
In Fig.~\ref{f:uni} we report the ratio
$\overline{\theta}/\theta^{*}$ and  variance
multiplied by ${H_M}$ [see Eq.~(\ref{VT:ineq})] as a function
of the number of measurements. As it apparent from the plots all
the curves approach one when the number of measurements increases. 
In the asymptotic region $H_M\simeq M G(\theta) \simeq M$. \\ 
 \par
Notice that the results here reported for $U_3(\theta)$ are actually valid for any other 
unitary gate. This can be seen as follows: any unitary
$U_{\boldsymbol n}(\theta)=\exp\left( -i \bmsigma_n \theta \right)$ describing rotations
around the arbitrary axis $n$ may be written as $U_{\boldsymbol n}= O U_3(\theta)
O^\dag$ where $O$ is the rotation corresponding to the mapping 
$3\rightarrow {\boldsymbol n}$. As a consequence, the optimal measurement 
corresponds to the projectors $\Pi_0^\prime = O^\dag |\Psi(0,\omega)\rangle\langle
\Psi(0,\omega)| O$, $\Pi_1^\prime= 1 - \Pi_0^\prime$ and the optimal probe is 
given by $O^\dag |0\rangle$.
\par
A question arises on whether the results reported above may be easily
implemented in practice. This concerns the stability of the measurement
rather than its precision. The point is the following: Suppose that for 
some reasons the values of parameters $\alpha$ (probe) $\beta$ (measurement)
slightly deviate from the optimal settings. To which extent 
the overall performances of the procedure are degraded? A convenient way
to address this issue is to make a perturbation analysis upon expanding 
the Fisher information $G(\theta)$ around the optimal settings 
$\alpha=\beta=\pi/2$ up to second order
\begin{align}
\label{expan}
G (\theta) & \simeq 1 -\frac1{\sin^2\theta} 
\left[(\alpha-\pi/2)^2+(\beta-\pi/2)^2\right]  + 2 \frac{\cos\theta}{\sin^2\theta}
(\alpha-\pi/2)(\beta-\pi/2)
\nonumber \\
& \stackrel{\theta\ll 1}{\simeq} 1 - \frac1{\theta^2}
(\alpha-\beta)^2\:.
\end{align}
\begin{figure}[ht]
\centerline{\includegraphics[width=0.8\textwidth]{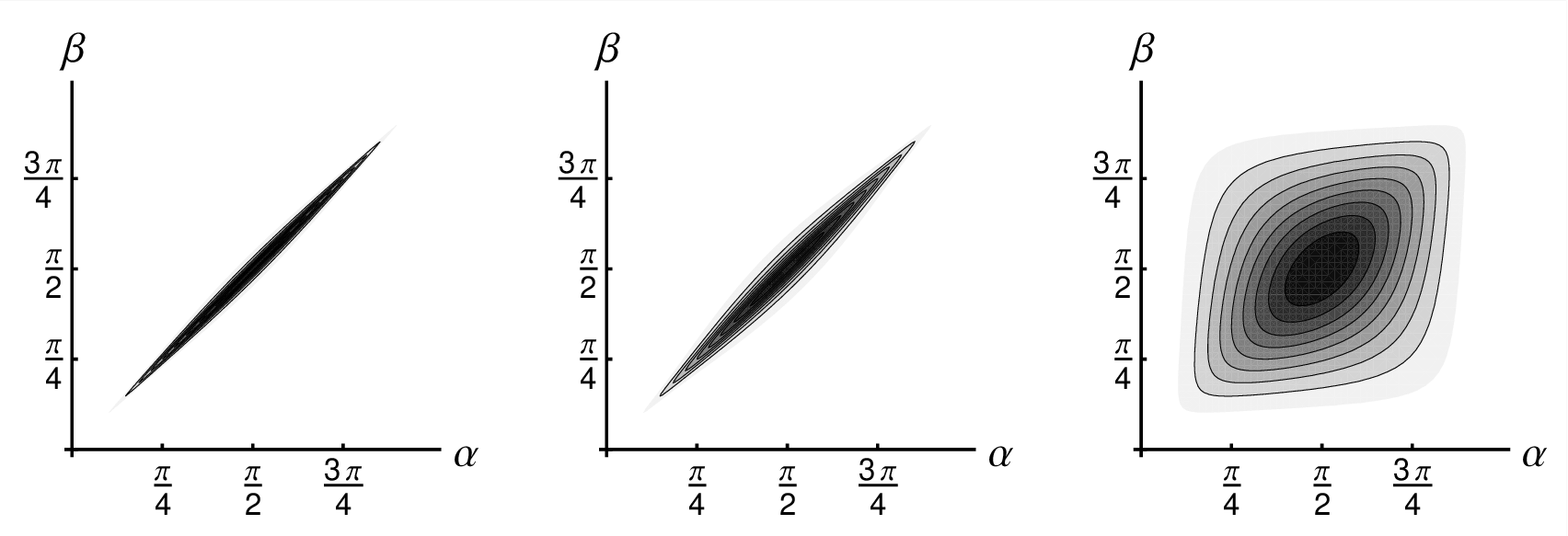}}
\caption{Contour plot of the Fisher information $G(\theta)$ as a function 
of the probe and measurement parameters $\alpha$ and $\beta$ for three 
different values of the gate parameter $\theta^*=0.05,0.1,1$. Darker
areas corresponds to higher values.}
\label{f:expan}
\end{figure} \par
Eq. (\ref{expan}) shows the quadratic decrease of $G$ out of the
optimal setting and, especially for small values of the gate parameter 
$\theta$, the dramatic effect of a mismatch between the values of $\alpha$ 
and $\beta$. The latter is well illustrated in Fig. \ref{f:expan} where
$G$ as a function of $\alpha$ and $\beta$ is shown for different values of
$\theta$. As it is apparent from the plots a mismatch
$|\alpha-\beta|\sim \theta$ of the order of the parameter to be
estimated is enough to make the whole procedure ineffective.
Fortunately, as we will see in Section \ref{ss:ent}, the stability issue may 
be overcome by using entangled probes and the optimal performances
still achieved by a two-qubit probe configuration.
\subsection{Monte Carlo simulated experiments}
\begin{figure}[h!]
\psfrag{P}{\tiny $P(\theta)$}
\includegraphics[width=0.5\textwidth]{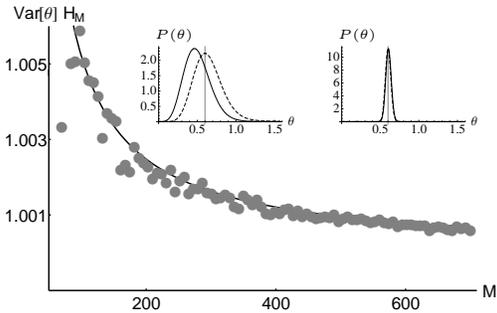}
\caption{Rescaled variance (variance multiplied by the generalized
Fisher information) of the a posteriori distribution as a function of the
number of measurements. Black line is for the asymptotic distribution
whereas gray squares are for the experimental one.
Inset: asymptotic (dashed line) and experimental 
(solid line) Bayesian a posteriori distributions 
for $\theta^{*}=0.6$ as obtained using optimal 
single-qubit probe and $M=20$ (left) or $M=500$ (right) repeated
measurements.} \label{f:sim}
\end{figure}

Before addressing stability of the measurement let us compare the 
asymptotic a {\em posteriori} distribution for the gate parameter with 
the results of Monte Carlo simulated experiments. This is in order 
to locate the asymptotic region and make quantitative statements on the
achievability of the ultimate bounds to precision. 
We have simulated $M$ repeated measurements using the optimal probe/measurement 
settings and inserted the resulting values of $m_j$, $j=0,1$ in Eq.
(\ref{ge}) to obtain the a posteriori distribution for the gate parameter.
In Fig. \ref{f:sim} we plot the rescaled variance of this distribution together
with the variance of the asymptotic a posteriori distribution of Eq.
(\ref{ga}). Remarkably the asymptotic region is achieved after a limited 
number of runs, thus proving that Bayesian approach may useful in practical 
applications. In the inset we report the full distribution, both the 
experimental a posteriori and the asymptotic one, for $M=20$ and $M=500$. 
Notice that: i) for $M=500$ the a posteriori experimental distribution 
is already indistinguishable from the asymptotic one and ii) the
asymptotic is already unbiased, {\em i.e} efficiency has been achieved 
for a limited number of runs.
\subsection{Estimation via two-qubit entangled probes}\label{ss:ent}
\begin{figure}[h!]
\includegraphics[width=0.6\textwidth]{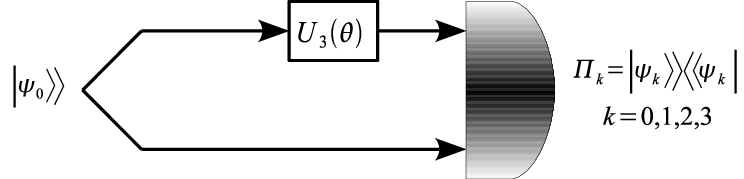}
\caption{Schematic diagram of parameter estimation by entangled qubit 
probe: $\lbrace|\psi_{0}\rangle\rangle\rbrace$ is prepared and is 
subjected to the unitary transformation $U_{\theta}$ on one qubit.
Finally, a projective measurement $\lbrace\Pi_{\alpha}\rbrace$ is 
performed.}\label{fig:bell} 
\end{figure}

An alternative scheme for gate estimation may be designed using entangled 
states as depicted in Fig.~\ref{fig:bell}. In fact, the use of entanglement 
may improve estimation \cite{Fujiwara,Matteo,gargpe,entdec}. In this section we 
investigate whether this is the case for the present estimation problem. 
Basically, the use of entanglement increases the dimension of Hilbert 
space and thus the number of possible outcomes of a measurement 
performed on the perturbed signal. The corresponding Fisher information 
does not increase but the maximum value is achieved for a large class
of probe signals. The Bayesian estimator is able to exploit this fact to 
increase the overall stability of the estimation
procedure of the gate parameter. 
\par
In order to see this behavior in practice let us consider the 
estimation of the parameter of the gate $U_3 (\theta)$ using a 
generic pure state of the form 
\begin{align}
|\psi_0\rangle\rangle = \frac{1}{\sqrt{2}}\sum_{k=0}^3 c_k |\sigma_k\rangle\rangle
\end{align}
where we used the matrix notation for states
$|A\rangle\rangle\doteq\sum_{ij}A_{ij}|i\rangle|j\rangle\equiv
A\otimes{\mathbbm I}|{\mathbbm I}\rangle\rangle$. 
As a measurement we consider the "Bell" measurement
made of the four projectors $\Pi_k=\frac12
|\sigma_k\rangle\rangle\langle\langle\sigma_k |$, $k=0,1,2,3$, 
over a set of maximally entangled states. 
After the evolution under the unitary $U_3(\theta)$ the four possible 
outcomes of  the measurements occur with the probabilities
\begin{align}
P_{0}(\theta)&= c_0^2 \cos^2\frac{\theta}{2} + c_3^2
\sin^2\frac{\theta}{2}\:, \quad 
P_{1}(\theta)= \left(c_1 \cos \frac{\theta}{2} - c_2 
\sin\frac{\theta}{2}\right)^2\:,
\nonumber\\
P_{3}(\theta)&= c_0^2 \sin^2\frac{\theta}{2} + c_3^2 \cos^2\frac{\theta}{2}
\:, \quad 
P_{2}(\theta)= \left(c_1 \sin \frac{\theta}{2} + c_2
\cos\frac{\theta}{2}\right)^2\:.
\end{align}
The corresponding Fisher information is given by
$$
G(\theta) = c_1^2 +c_2^2 + \frac{(c_0^2-c_3^2)(c_0^4-c_3^4)\sin^2
\theta}{(c_0^2+c_3^2)-(c_0^2-c_3^2)\cos^2\theta}\, $$
and achieves its maximum ($G=1$, $\forall \theta$) for any state 
with $c_0=0$ or $c_3=0$. Therefore, having fixed the Bell measurement at
the output, a possible deviation in the probe preparations is not
degrading the performances of the estimation procedure. 
\section{Conclusions}
\label{s:remarks}
In this paper we have analyzed estimation of one-parameter unitary gates
for qubit systems. We have addressed Bayesian estimation procedures and
compared their performances with the ultimate quantum limits to
precision.  Bayes estimator is known to be asymptotically unbiased, but
for practical implementation is of interest to evaluate quantitatively
how many measurements are needed to achieve the asymptotic region.  To
this aim, after the evaluation of the asymptotic  {\em a posteriori}
distribution for the gate parameter, we have compared it to the
distribution obtained by Monte Carlo simulated experiments and full
Bayesian analysis and shown that asymptotic optimality of Bayes
estimator is achieved after a limited number of runs.  We have also
addressed the issue of stability, {\em i.e} the robustness of the
optimal settings against fluctuations of the probe and measurement
parameters. It has been shown that the use of entanglement is useful to
improve stability. More explicitly, we have shown that, although the
Fisher  information  does not increase, its maximum value is achieved
for a large class of probe signals, thus making the procedure more
robust and increasing the overall stability of the estimation procedure.
\section*{Acknowledgments}
The authors thank M. Borrelli, M. Zaro, and N. Tomassoni for useful
discussions. This work has been partially supported by the CNR-CNISM
convention. This article was completed at a time of drastic cuts to
research budgets imposed by the Italian government \cite{nn08}; as a 
result research is becoming increasingly difficult in Italian universities 
and may in the near future be brought to a complete halt.

\end{document}